\documentclass[sigplan,10pt,screen,nonacm]{acmart}
\renewcommand\footnotetextcopyrightpermission[1]{}

\usepackage{booktabs}   
\usepackage{subcaption} 
\usepackage{graphicx}
\usepackage{multirow}
\usepackage{algorithmic} 
\usepackage{algorithm}
\usepackage{xspace}
\usepackage{siunitx}
\usepackage{microtype}
\usepackage{tikz}
\usepackage{pgf-umlsd}

\newcommand{\rawname}{MemExchange}
\newcommand{\name}{\textit{\rawname}\xspace}

\usepackage[normalem]{ulem}

\begin{document}

\title[\rawname: Utility-Driven Distributed Memory Reallocation for Multi-Tenant Datacenters]{\rawname: Utility-Driven Distributed Memory Reallocation for Multi-Tenant Datacenters}                              


\author{AmirHossein Seyri}
\affiliation{
  \institution{University of Illinois Chicago}            
 \city{Chicago}
 \state{IL}
  \country{USA}                    
}
\email{amirhossein.seyri@gmail.com}          

\author{Abhisek Pan}
\affiliation{
  \institution{Microsoft}           
 \city{Redmond}
 \state{WA}
  \country{USA}                   
}
\email{abpan@microsoft.com}         

\author{Balajee Vamanan}
\affiliation{
  \institution{University of Illinois Chicago}            
 \city{Chicago}
 \state{IL}
  \country{USA}                    
}
\email{bvamanan@uic.edu}          


\begin{abstract}

To handle unpredictable workloads, cloud providers typically over-provision memory to meet peak demand, resulting in substantial underutilization across datacenter clusters. At the same time, memory-constrained tenants may suffer elevated cache miss rates, even when idle capacity remains stranded elsewhere in the infrastructure.
MemExchange is a cluster-wide, multi-tenant memory management system that dynamically right-sizes in-memory caching tenants according to workload demand. Leveraging marginal-utility–based allocation derived from online Miss Ratio Curve (MRC) estimation, MemExchange redistributes idle memory between tenants across physical nodes using RDMA. This approach transforms the dedicated caching memory scattered across servers into a logically aggregated pool, enabling cross-node memory exchange without centralized coordination or forced tenant co-location. To support efficient remote access, we design the MemExchange Tracker Communication (MTC) protocol, an application-layer mechanism that coordinates memory reallocation and enables one-sided RDMA operations without involving remote CPUs.
We implement MemExchange in Memcached and evaluate it through microbenchmarks, medium and rack-scale deployments of up to 100 CloudLab servers. Our results show up to $2.3\times$ lower remote-access overhead compared to TCP-based designs, a 13\% increase in cluster-wide memory utilization at rack scale, and up to 63\% reduction in miss rate for memory-constrained tenants under skewed workloads. 

\end{abstract}



\maketitle

\section{Introduction}
\label{sec:intro}

Memory caches and key-value stores are essential components of modern web services. By storing frequently accessed objects in DRAM, they reduce backend database load and improve request latency by orders of magnitude. A cache hit is served directly from memory, whereas a cache miss requires retrieving the item from a backend database, often multiple network hops away and backed by storage significantly slower than DRAM. Even small improvements in cache hit rate can have tremendous impact; for example, a 1\% increase in hit rate can reduce request latency by 35\%~\cite{cidon2016cliffhanger}. Maintaining high hit rates is therefore critical for user-facing services.

Achieving high hit rates requires the cache’s working set to fit within its allocated memory. However, working set sizes are often dynamic and may exceed the statically provisioned memory assigned to a tenant~\cite{wu2022nyxcache, cidon2015dynacache, cidon2016cliffhanger, seyri2022memsweeper, seyri2019dynamically}. Static allocation and manual right-sizing are inefficient for such workloads and frequently lead to over-provisioning. As a result, some tenants suffer from memory pressure and elevated miss rates, while others hold idle memory.

This imbalance is amplified in multi-tenant cloud environments. Commercial cloud providers such as Microsoft Azure and AWS offer caching as a service (e.g., Redis-based offerings~\cite{azure:red, redis:xxx}). Multiple tenants are co-located on shared infrastructure~\cite{hadary2020protean, wu2022nyxcache}, which improves consolidation but leads to rigid per-tenant memory quotas. Despite extensive work on multi-tenant cache management~\cite{cidon2016cliffhanger, memshare203165, pu2016fairride, seyri2019dynamically, seyri2022memsweeper}, most prior systems operate within a single physical server and cannot reclaim idle memory across the cluster.

Meanwhile, cluster-wide memory utilization in large-scale datacenters often remains between 40\% and 60\%~\cite{maruf2023memtrade}, leaving substantial idle capacity distributed across thousands of servers. This underutilization stems primarily from conservative over-provisioning: tenants are allocated memory based on peak demand to satisfy Service Level Objectives (SLOs), even though their steady-state working sets are frequently much smaller. Because these allocations are enforced per server and per tenant, idle memory on one node cannot be repurposed to relieve memory pressure elsewhere in the cluster. As a result, some tenants experience elevated miss rates due to insufficient memory, while unused capacity remains stranded across the infrastructure. If this fragmented idle memory could be dynamically and safely redistributed to memory-constrained tenants, both overall memory utilization and per-tenant cache hit rates could improve simultaneously.

Enabling such redistribution across physical nodes requires efficient remote memory access. Remote Direct Memory Access (RDMA) provides low-latency, high-throughput communication that bypasses the remote CPU and operating system~\cite{dragojevic2014farm}. Although remote memory access is slower than local DRAM, the overhead is modest. In our evaluation, a workload in which 94\% of GET requests were served from remote memory incurred an average latency of 31\,$\mu$s compared to 22\,$\mu$s for purely local access—an additional cost of approximately 9\,$\mu$s per request. This overhead is negligible relative to the latency of a cache miss that requires fetching data from a backend database (e.g., $\sim$10\,ms for MySQL~\cite{cidon2016cliffhanger}). Thus, even when most requests are served remotely, the latency increase is orders of magnitude smaller than the penalty avoided by preventing a miss.

In this paper, we present \textbf{MemExchange}, a cluster-wide multi-tenant memory management system that dynamically right-sizes caching tenants by reallocating idle memory across physical nodes. MemExchange extends marginal-utility–based resizing beyond a single machine, enabling memory to be transferred between any two tenants in the cluster. Using RDMA, remote memory becomes a transparent overflow tier for under-provisioned tenants while preserving fast local access for hot data.

MemExchange unifies the dedicated caching memory scattered across hundreds of servers into a logically aggregated pool. Rather than relying solely on careful VM placement or bin-packing strategies, MemExchange dynamically reallocates capacity according to workload demand. A distributed \emph{tracker} on each server maintains metadata about available memory, tenant allocations, and reallocation decisions. Trackers coordinate through the \emph{MemExchange Tracker Communication (MTC)} protocol, enabling fully distributed memory exchange without centralized brokers.

\subsection{Contributions.}
The main contributions of this paper are:

\begin{itemize}\setlength{\itemsep}{0pt}
    \item \textbf{Cluster-wide utility-aware resizing:} We extend marginal-utility–based cache resizing to operate across physical servers, enabling dynamic redistribution of memory between tenants throughout the cluster.
    
    \item \textbf{RDMA-enabled remote caching tier:} We integrate RDMA operations directly into Memcached to support low-latency remote memory access without involving remote CPUs.
    
    \item \textbf{Fully distributed coordination:} We design the MemExchange Tracker Communication (MTC) protocol, an application-layer mechanism for decentralized memory reallocation.
    
    \item \textbf{Comprehensive cloud-scale evaluation:} We evaluate MemExchange on up to 100 CloudLab servers~\cite{Duplyakin+:ATC19}, demonstrating up to $2.3\times$ lower remote-access overhead than TCP-based designs, a 50\% utilization increase in medium-scale benchmarks, a 13\% increase at rack scale, and up to 63\% miss-rate reduction under skewed workloads.
\end{itemize}

\section{Background}
\label{sec:background}

This section reviews the background underlying MemExchange, including Memcached’s memory organization, MRC-based resizing with shadow queues and marginal utility, RDMA, and the memory underutilization that motivates cluster-wide reallocation.


Memcached~\cite{Mem:xxx} uses a slab-based allocator that divides memory into fixed-size \SI{1}{MB} pages, each assigned to a slab class for a predefined item size. Pages are partitioned into equal-sized chunks, and items are placed into the smallest class that can accommodate them. Each slab class maintains a segmented LRU (hot, warm, and cold). Frequently accessed items are promoted toward the hot segment, while colder items migrate toward the cold segment and become eviction candidates. Although this design reduces memory fragmentation, each instance operates with a fixed memory quota. In multi-tenant deployments, idle memory in one tenant cannot be reallocated to others, leading to cluster-wide inefficiencies.

\subsection{Miss Ratio Curves and Shadow Queues}
\label{sec:MRC}

Dynamic memory resizing requires estimating how additional memory would affect cache performance. Miss Ratio Curves (MRCs) characterize the relationship between cache size and miss rate \cite{mattson1970evaluation}. Traditional MRC construction relies on stack distance or reuse distance analysis \cite{ding2003predicting, olken1981efficient}, which can be computationally expensive.

MemExchange leverages shadow queues \cite{cidon2016cliffhanger, seyri2019dynamically} to approximate MRCs online with minimal overhead. When an item is evicted from the main LRU, its shadow remains in a secondary queue; its value is removed but its metadata (e.g., key and flags) is retained. Because shadow entries omit object payloads, they are lightweight and can be stored efficiently.

Each shadow page mirrors the structure of its corresponding slab class and holds the same number of entries as a regular page (e.g., 13,107 items per page in class 1~\cite{Mem2:xxx}). When a request misses in the main cache but hits in the shadow queue, this indicates that the item would have been a hit if the cache were larger. The position of the shadow entry in the LRU provides an estimate of the additional memory required to prevent similar misses in the future. By tracking shadow hits across different reuse-distance intervals, tenants approximate the local slope of their MRC, quantifying the expected hit-rate improvement per additional page of memory.

\subsection{Marginal Utility–Driven Allocation}
\label{sec:MarginalUtility}

A sustained concentration of shadow hits within a reuse-distance interval signals unmet demand, and the position of that interval determines the additional memory requested by the tenant. Because tenant quotas and server capacity are finite, increasing one tenant’s allocation requires reclaiming memory from another, ideally from tenants whose hit rates will degrade the least.

To identify \emph{victors} (tenants seeking memory) and \emph{victims} (tenants that can relinquish memory), we adopt the scoring method of~\cite{seyri2022memsweeper}, based on Marginal Utility (MU)~\cite{qureshi2006utility}. MU quantifies the expected hit-rate improvement per additional page of memory. A high MU indicates strong benefit from additional capacity, while a low/zero MU indicates surplus memory that can be reclaimed with minimal impact. Tenants compute MU locally using internal statistics (e.g., hit counters on shadow/main pages) but report only their scores to the tracker. Internal cache state remains private. When reallocation occurs, the victim releases its lowest-utility page, minimizing performance degradation.

This utility-driven mechanism enables principled, workload-aware resizing of tennats. 
Unlike prior intra-server approaches, MemExchange generalizes this utility-driven mechanism to operate across the entire cluster, enabling memory redistribution between tenants regardless of their physical placement. When tenants reside on different nodes, memory is reallocated and accessed via RDMA.

\subsection{RDMA Fundamentals}

Remote Direct Memory Access (RDMA) enables direct memory-to-memory transfers between machines without involving the remote CPU or operating system. By bypassing the TCP/IP stack and kernel, RDMA provides lower latency and higher throughput than traditional networking in datacenters.

Applications interact with the specialized NIC using RDMA verbs, including one-sided operations such as \texttt{RDMA\_READ} and \texttt{RDMA\_WRITE}, and two-sided operations such as \texttt{Send}/\texttt{Receive}. Operations are issued through a Queue Pair (QP) and completion events are reported via a Completion Queue (CQ). One-sided operations access remote memory without remote CPU intervention, while two-sided operations require coordination at both endpoints. To enable remote access, memory regions must be registered with the NIC~\cite{rdmamojo:mojo}, producing a Memory Region (MR) and associated \emph{lkey} and \emph{rkey}. Before issuing one-sided operations, endpoints exchange metadata (e.g., address and rkey) via an initial two-sided operation.

RDMA operations copy data between registered memory regions without guaranteeing application-level correctness. For \texttt{RDMA\_WRITE}, data must be prepared locally and the buffer cannot be reused until completion. For \texttt{RDMA\_READ}, the local destination must not be accessed until the operation completes. In multi-threaded systems such as Memcached, concurrent RDMA operations must use non-overlapping buffers to avoid corruption. Because memory accounts for a significant fraction of server cost~\cite{fuerst2022memory}, efficient buffer management is critical. MemExchange uses a small \SI{2}{MB} registered region (split between reads and writes) to safely support thousands of concurrent RDMA operations while minimizing memory overhead.

\subsection{Memory Underutilization in Clouds Environments}

Major cloud providers (Azure, AWS, Google Cloud) offer flexible resources enabling rapid adaptation without significant hardware investment. Emerging applications like machine learning and generative AI (e.g., Large Language Models) have sharply increased demand for cloud resources. Although resource management has been widely studied~\cite{younge2010efficient, cortez2017resource, perez2022dynamic, gu2017efficient, li2023pond, fuerst2022memory}, memory utilization remains notably low —around 24\% in Alibaba's HPC systems~\cite{panwar2019quantifying}, 42\%-65\% in its datacenters~\cite{jiang2019characteristics, chen2018does}, and below 60\% in Google Cloud~\cite{maruf2023memtrade}.

This inefficiency arises primarily from over-provisioning and dynamic workloads. Tenants are typically allocated memory based on peak demand to satisfy Service Level Objectives (SLOs), leaving substantial idle capacity during operation. These observations motivate a cluster-wide mechanism capable of dynamically redistributing memory according to workload demand while preserving performance isolation and minimizing latency overhead.
\section{Related Work}
\label{sec:related}

Prior research has addressed memory efficiency in datacenters through cache resizing, memory harvesting, remote memory systems, and memory disaggregation. While these systems improve either utilization or performance, none jointly support cluster-wide dynamic resizing with marginal-utility–driven allocation and RDMA-enabled remote memory in a fully distributed architecture.

\subsection{Cache Performance and MRC Estimation}

MemExchange builds on a long line of work in cache analysis and performance modeling. Mattson et al.~\cite{mattson1970evaluation} introduced the concept of stack distance and demonstrated how it can be used to construct Miss Ratio Curves (MRCs). Efficient reuse-distance tracking mechanisms were later proposed in \cite{olken1981efficient, ding2003predicting}, forming the basis for online MRC approximation. Shadow queues and their use in approximating MRCs were studied in systems such as Cliffhanger~\cite{cidon2016cliffhanger} and MemShare~\cite{memshare203165}. The balanced search tree techniques (e.g., AVL trees) used to track reuse distance and compute marginal benefit in MemExchange follow the approach of \cite{seyri2022memsweeper}. The concept of \emph{Marginal Utility} (MU) and the LookAhead algorithm were introduced in \cite{qureshi2006utility}, which formalized utility-driven resource allocation. While these works established the theoretical and algorithmic foundation for cache resizing and utility-aware allocation, they primarily operate at the single-tenant or single-server level. MemExchange extends these principles to cluster-wide memory reallocation.

\subsection{Memory Management and Autoscaling}

Managing the memory of caching tenants has been widely studied. Prior work optimizes eviction and allocation policies within single-tenant caches~\cite{beckmann2015talus,cidon2016cliffhanger,beckmann2018lhd,hu2015lama,yang2021segcache,nishtala2013scaling}. Multi-tenant systems address allocation among co-located tenants on the same server~\cite{cidon2015dynacache,memshare203165,wu2022nyxcache,seyri2022memsweeper,seyri2019dynamically,byrne2018mpart,berger2018robinhood,pu2016fairride}. MemSweeper~\cite{seyri2022memsweeper} is particularly relevant. It dynamically adjusts memory allocation among tenants based on estimated Miss Ratio Curves derived from shadow queues and reuse-distance tracking. 
However, these systems are confined to server-level allocation. They cannot reclaim or redistribute memory across physical nodes, leaving cluster-wide idle capacity unexploited. MemExchange adopts a similar MRC-based utility framework but extends resizing across the entire cluster, enabling inter-node memory reallocation.

\subsection{Memory Harvesting}

Memory harvesting mechanisms reclaim unused memory from one application, VM, or server and reallocate it to others with higher demand, improving overall utilization.

Memtrade~\cite{maruf2023memtrade} enables public cloud VMs (\emph{producers}) to offer memory to remote VMs (\emph{consumers}) through a centralized broker. Consumers submit memory requests to the broker, while producers periodically report available capacity. The broker coordinates memory reclamation and monitors performance via swap-in events, returning memory when degradation is detected. Memtrade relies on key-value stores such as Redis~\cite{redis:xxx}, requiring a dedicated Redis instance per producer-consumer pair. While previously allocated remote memory remains accessible if the broker fails, any broker crash/unavailability halts future remote reallocations cluster-wide.

In contrast, MemExchange employs a fully distributed tracker architecture, confining failures to individual nodes and preventing cluster-wide reallocation shutdowns (see \S\ref{sec:fault-tolerance}). Moreover, Memtrade emphasizes resource matching and tenant-driven participation using price-per-hit values derived from estimated MRCs, but its methodology for constructing these curves is not clearly specified. MemExchange instead builds MRCs using the approach of \cite{seyri2022memsweeper}, which has been shown to outperform techniques such as Cliffhanger~\cite{cidon2016cliffhanger, seyri2022memsweeper}. Finally, Memtrade exposes remote memory through a KV-based interface. MemExchange leverages RDMA for direct memory access, yielding significantly lower remote access latency, as demonstrated in Microbenchmark A (\S\ref{sec:rdmavstcp}).

Hypervisor-level ballooning approaches~\cite{waldspurger2002memory, fuerst2022memory} reclaim memory within a single host but operate below the application layer.


\subsection{Remote Memory and Disaggregation}

Memory disaggregation decouples memory from compute nodes and pools it into a network-accessible resource shared across tenants. High-speed interconnects such as RDMA and CXL~\cite{CXL:xxx, li2023pond} make such architectures feasible.

INFINISWAP~\cite{gu2017efficient} uses RDMA to swap memory pages to remote idle nodes instead of local disk. Its decentralized design improves swap latency and increases effective memory capacity. However, INFINISWAP does not reclaim idle memory from co-located tenants and assumes tenants are correctly sized. It supplements local memory with remote swap even when local redistribution may be more efficient. MemExchange instead dynamically right-sizes tenants based on workload demand and reallocates memory cluster-wide according to marginal utility.

FaRM~\cite{dragojevic2014farm} provides a distributed memory platform in which applications share a cluster-wide address space. It uses RDMA to access remote objects and manages memory in \SI{1}{MB} slabs similar to Memcached. While FaRM emphasizes high-throughput transactional access to distributed memory, it does not perform workload-aware resizing or inter-tenant memory reallocation.

HERD~\cite{kalia2014using} and MICA~\cite{lim2014mica} focus on minimizing request latency using RDMA write-based request injection and server-side polling. Redy~\cite{zhang2021redy} and GAM~\cite{cai2018efficient} similarly leverage RDMA verbs and unified memory abstractions to maximize throughput and reduce latency compared to TCP/IP-based systems. These systems optimize request processing performance but do not incorporate marginal-utility–based resizing or cluster-wide memory redistribution.


\paragraph{Summary} Prior work addresses cache modeling, intra-server resizing, memory harvesting, and remote-memory architectures, but treats these dimensions in isolation. MemExchange combines cluster-wide dynamic resizing, marginal-utility–based allocation via shadow queues, fully distributed coordination, and RDMA-enabled remote memory. Our evaluation therefore compares against MemSweeper~\cite{seyri2022memsweeper} and INFINISWAP~\cite{gu2017efficient}, which represent the closest approaches along the resizing and remote-memory dimensions.

\section{\name's Design}
\label{sec:design}

MemExchange extends Memcached with a distributed, utility-driven memory reallocation framework that enables cluster-wide dynamic resizing without centralized control.

\subsection{Architecture Overview}

On each physical server, MemExchange consists of multiple tenants and a lightweight process called the \emph{Tracker}. The Tracker oversees a shared-memory region from which all local tenants draw their memory. It maintains metadata describing the total shared-memory size, allocated and unallocated regions, and per-tenant limits. In addition, it coordinates memory reallocations both locally and across servers, and facilitates RDMA metadata exchange between remote tenants.

The shared-memory region is initialized and managed by the Tracker at server startup. Tenants allocate memory using \texttt{mmap} and release it using \texttt{munmap}, ensuring that all memory requests are mediated through the shared-memory region rather than through standard heap allocation. This design allows the Tracker to maintain global visibility over server-level memory availability while preserving process isolation between tenants.

Each tenant is initially assigned memory equal to its purchased capacity. However, these allocations are not strictly static. Because the total server-level caching memory is finite, and because cluster-wide memory utilization is often significantly below peak capacity~\cite{maruf2023memtrade}, MemExchange treats memory as a dynamically adjustable resource. Tenants may temporarily exceed their purchased capacity if the reallocation mechanism determines that doing so increases overall memory utility. Conversely, tenants with surplus capacity may relinquish memory with minimal impact on their hit rate.

MemExchange preserves both performance and security isolation. From a performance standpoint, a tenant operating within MemExchange performs at least as well as an isolated Memcached instance without memory sharing~(\S\ref{sec:mediumscalebenchmarks}). From a security standpoint, standard kernel page protections prevent unauthorized cross-tenant access. Pages selected for reallocation are cleared and their contents safely evicted before reuse. When pages are exposed as remote memory regions, access is granted explicitly through RDMA registration, ensuring that only authorized tenants can access them.


\subsection{MemExchange Tracker Communication (MTC)}

The MemExchange Tracker Communication (MTC) protocol enables distributed coordination of memory reallocations across the cluster without relying on a centralized controller. Because tenants may reside on different physical servers, and because the reallocation mechanism requires identifying the most suitable victim tenant cluster-wide, Trackers must exchange lightweight coordination messages to discover and finalize victim–victor pairings.

When a tenant’s score, as determined by the reallocation mechanism, exceeds the currently known highest score in the cluster, the Tracker managing that tenant initiates the MTC protocol. The initiating Tracker multicasts a small request message to all other Trackers in the cluster. This message announces the presence of a potential victor and includes its current score. Upon receiving this request, each remote Tracker evaluates its locally managed tenants and responds via unicast with the identity and score of its lowest-scored candidate that could serve as a victim.

The initiating Tracker collects responses within a bounded discovery window. After this window expires, it selects the most suitable victim among the received candidates (the tenant with the lowest score) and proceeds to notify the corresponding Tracker of its intent to reallocate memory. This notification constitutes the second phase of the protocol, during which the victim’s Tracker acknowledges the request and prepares the selected page for release. To prevent ambiguity between overlapping requests, each message is assigned a unique Message Identifier (MID). All responses include this identifier, allowing the initiating Tracker to discard stale or unrelated replies. 

MTC uses small UDP packets for inter-Tracker communication. Because the protocol does not require strict reliability guarantees, lost or delayed packets do not introduce deadlock or inconsistency. If no suitable responses are received within the discovery window, or if direct communication with a selected victim fails, the initiating Tracker simply times out and restarts the protocol. This design favors simplicity and scalability over transport-layer reliability, which would impose additional overhead in large clusters.

Once a victim is selected, the protocol transitions to RDMA setup. If the victor and victim tenants reside on different servers, the victim registers the selected page as a remote memory region and transmits the required metadata (specifically the remote address and rkey) using a two-sided \texttt{RDMA\_SEND} operation. Before this exchange occurs, both tenants determine whether an RDMA connection between them already exists. Because establishing RDMA connections is relatively expensive, MemExchange maintains persistent connections once established. Each tenant keeps a table of active connections indexed by the tuple (IP address, Memcached port), which uniquely identifies a tenant within the cluster. If no connection exists, one is established prior to metadata exchange; otherwise, the existing connection is reused.

After metadata exchange completes, subsequent accesses to the reallocated page proceed via one-sided RDMA operations (\texttt{RDMA\_READ} and \texttt{RDMA\_WRITE}), allowing the victor to access remote memory without involving the remote CPU. The completion of this phase marks the end of the MTC protocol for that page transfer.

\begin{figure}[]
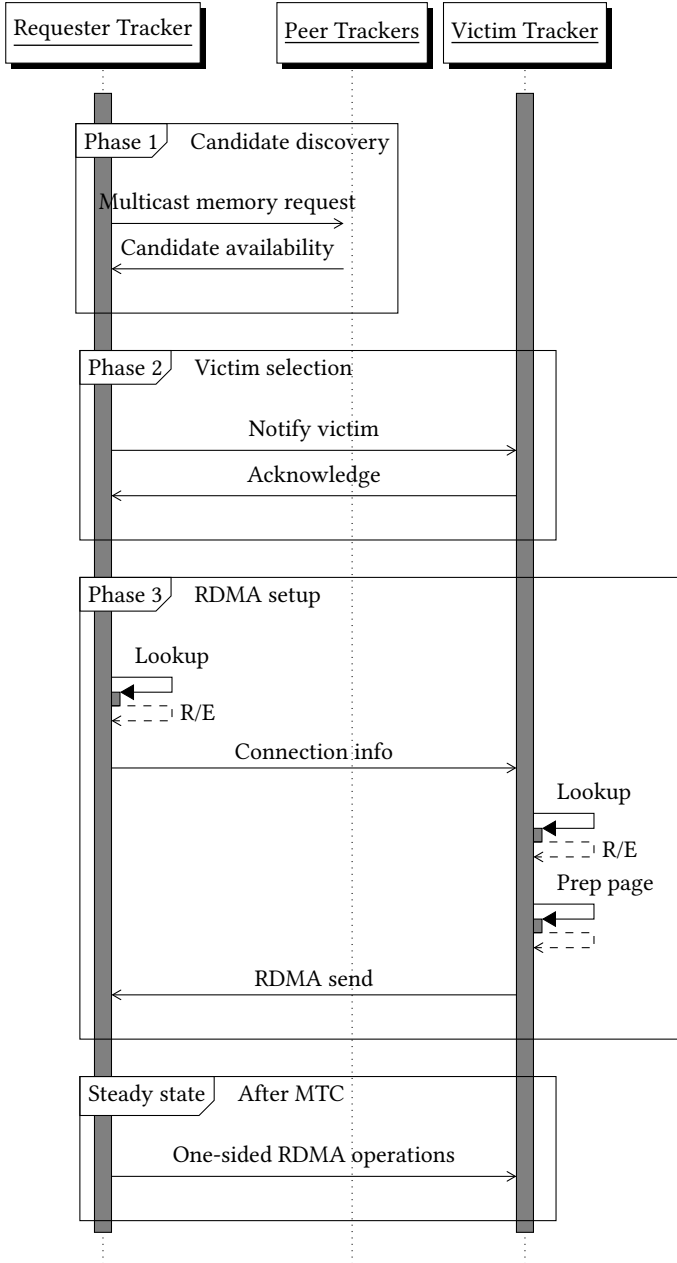

\centering
\small
\begin{sequencediagram}

\newthread[gray]{r}{Requester Tracker}
\newinst[1.0]{p}{Peer Trackers}
\newthread[gray]{v}{Victim Tracker}

\begin{sdblock}{Phase 1}{Candidate discovery}
  \mess{r}{Multicast memory request}{p}
  \mess{p}{Candidate availability}{r}
\end{sdblock}

\begin{sdblock}{Phase 2}{Victim selection}
  \mess{r}{Notify victim}{v}
  \mess{v}{Acknowledge}{r}
\end{sdblock}

\begin{sdblock}{Phase 3}{RDMA setup}
  \begin{callself}{r}{Lookup}{R/E}
  \end{callself}

  \mess{r}{Connection info}{v}

  \begin{callself}{v}{Lookup}{R/E}
  \end{callself}

  \begin{callself}{v}{Prep page}{}
  \end{callself}

  \mess{v}{RDMA send}{r}
\end{sdblock}

\begin{sdblock}{Steady state}{After MTC}
  \mess{r}{One-sided RDMA operations}{v}
\end{sdblock}

\end{sequencediagram}
\vspace{-0.6em}
\caption{MTC protocol in MemExchange. R/E denotes reuse or establishment of RDMA connections.} 
\label{mtc_protocol}
\end{figure}

Figure~\ref{mtc_protocol} illustrates the sequence of interactions in the MTC protocol. Next, we discuss the reallocation mechanism that governs victim–victor selection and page transfers.

\subsection{Reallocation Mechanism}

The reallocation mechanism determines when, how much, and between which tenants memory is transferred. It implements workload-aware resizing by continuously evaluating each tenant’s memory demand and identifying opportunities to improve cluster-wide utility through page exchanges.

Each tenant periodically computes two scalar scores that estimate its expected benefit from acquiring an additional memory page or, conversely, the expected loss from relinquishing one. The score is grounded in marginal utility, which captures the change in expected cache hits per additional/fewer page of memory, and incorporates workload-sensitive signals such as recent miss rate, reuse-distance observations from the shadow queue, current allocation size, and historical memory transfers.

Let $S_0$ denote the extreme marginal utility of a tenant’s queues, specifically, the maximum marginal utility over the shadow queues and the minimum marginal utility over the main queues. The final score is computed as (for both queues):

\begin{equation}
S_1 = S_0 \cdot \left(\frac{N_r}{N_g}\right) \cdot MR_{ins} \cdot t \cdot \frac{1}{M}
\end{equation}

where $N_r$ and $N_g$ are the cumulative numbers of pages released and gained, $MR_{ins}$ captures recent per-second miss rate, $t$ encodes temporal recency, and $M$ normalizes by current allocation. A high shadow queue score indicates that a tenant is memory-constrained and likely to benefit from additional capacity (a potential \emph{victor}), whereas a low main queue score indicates that a tenant can relinquish memory with minimal impact (a potential \emph{victim}). Tenants compute these scores independently and report only the final values to their local Tracker, preserving encapsulation and minimizing coordination overhead.

At any given time, the system prioritizes the tenant with the highest score in the cluster. When a local tenant’s score exceeds the currently known global maximum, its Tracker initiates the MTC protocol to discover suitable victims. Trackers do not propagate every score update. Instead, a Tracker initiates coordination only when one of its local tenants exceeds the currently known global maximum score. This event-driven propagation ensures that global state is updated only when necessary, avoiding unnecessary coordination overhead while still maintaining an accurate view of the highest-scored tenant in the cluster. If a new tenant surpasses the current victor during an ongoing reallocation sequence, the previous victor transitions into a waiting state after completing its current page transfer. This ensures that reallocations proceed in descending order of expected utility gain, preserving the objective of maximizing cluster-wide benefit.

Victim discovery occurs during the first phase of MTC. The initiating Tracker aggregates candidate responses from peers and selects the tenant with the lowest reported score. However, selection is not purely score-driven. MemExchange incorporates locality-awareness into victim selection to reduce the performance overhead of remote access. Local tenants are preferred over remote ones whenever their scores are comparable. For remote candidates, physical proximity implicitly influences selection due to bounded discovery windows: Trackers that respond earlier are more likely to be considered. As a result, victims tend to be chosen from the same server or from nearby servers, minimizing additional latency introduced by RDMA operations. This design balances utility maximization with practical performance considerations, ensuring that reallocations do not introduce unnecessary cross-cluster latency.

Memory is transferred one page at a time. Even when a victor requests multiple pages, reallocations are performed sequentially. Because tenant scores evolve dynamically as workloads shift and pages are gained or released, selecting multiple victims in advance could result in suboptimal or outdated decisions. By reevaluating the system state between page transfers, MemExchange ensures that each reallocation reflects the most current demand conditions.

Page transfers are atomic. Once a victim is selected and a page reallocation begins, the transfer proceeds to completion w/o interruption. Victor transitions and MTC initiations by a newly emerged higher-scored tenant occur only between page transfers. This guarantees consistency and prevents partially executed reallocations.

The page chosen for release by a victim is always its lowest-utility page, ensuring minimal degradation in hit rate. Over time, as workloads evolve, tenants may alternate between being victors and victims, reflecting shifting demand. This continuous score-driven adaptation enables MemExchange to dynamically balance memory across the cluster without centralized coordination or static provisioning decisions.

\subsection{Remote Memory Access}

MemExchange extends Memcached with transparent support for remote memory, enabling tenants to utilize capacity located on other physical servers while preserving the semantics of a local cache. Once a page has been reallocated through the reallocation mechanism and coordinated via MTC, it becomes accessible to the victor tenant through RDMA. From the tenant’s perspective, this page augments its effective cache capacity, even though it physically resides in another server’s memory.

Unlike local reallocation, remote reallocation does not transfer ownership of the underlying physical memory. The victim tenant retains process-level ownership of the page but registers it as an RDMA memory region, granting controlled access permissions to the victor. Because RDMA registration pins the memory region, the operating system cannot swap it out, ensuring that remote pages remain resident in DRAM and suitable for high-performance access.

To integrate remote pages into Memcached without modifying its core memory layout assumptions, MemExchange introduces an abstraction called a \emph{remote item}. Remote items are lightweight local structures that represent objects stored in remote memory. Each remote item contains the key, metadata fields required for cache management, and the remote address and \texttt{rkey} identifying the corresponding memory region on the victim’s server. The value itself is not stored locally; it resides exclusively in remote memory.

This indirection allows the tenant to track remote objects using standard cache mechanisms while delegating value storage to remote pages. When a request references a remote item, the tenant retrieves the associated remote address and \texttt{rkey} and issues the appropriate one-sided RDMA operation to access the value. In this way, remote objects participate in normal cache operations without requiring the remote CPU to process requests.

\subsubsection{Victim-Side Responsibilities}

When instructed by its Tracker to relinquish a page, the victim tenant first selects its lowest-utility page. All cached items in that page are safely evicted, and the page contents are cleared to prevent residual data exposure. The page is then registered as an RDMA memory region with the necessary access permissions.

Because one-sided RDMA operations require the remote address and \texttt{rkey}, these metadata must be communicated to the victor. This exchange is performed using a two-sided \texttt{RDMA\_SEND} operation, which consumes a pre-posted receive request on the victor’s queue pair. Once this metadata transfer completes, the page becomes accessible via one-sided \texttt{RDMA\_READ} and \texttt{RDMA\_WRITE} operations.

Importantly, after the initial \texttt{RDMA\_SEND}, the victim does not participate in subsequent data accesses to that page. Remote reads and writes occur without remote (victim) CPU involvement.

\subsubsection{Victor-Side Responsibilities}

Upon receiving the metadata for a newly reallocated page, the victor integrates it into its slab class as remote capacity. Unlike locally allocated pages, remote pages cannot be directly mapped into the victor’s address space. Instead, the victor initializes a local representation of empty slots corresponding to that slab class and prepares the page layout remotely using an \texttt{RDMA\_WRITE}. This ensures that the remote page mirrors the structure expected by Memcached for that slab class.

Subsequent \emph{SET} operations may store objects in remote slots when local capacity is exhausted. Because \texttt{RDMA\_WRITE} operations can be issued asynchronously, remote insertions do not block request handling. In contrast, \emph{GET} operations require synchronous \texttt{RDMA\_READ} operations, as the value must be retrieved before responding to the client. Despite this, empirical results show that the additional latency introduced by RDMA is small compared to backend miss penalties (Section~\ref{sec:microA}).

\subsubsection{Modified Insertion and Lookup Paths}

MemExchange modifies Memcached’s insertion and lookup paths to incorporate remote memory as a secondary storage tier.

For \emph{SET} operations, the system first attempts to allocate space within local pages of the corresponding slab class. If no local slots are available and local allocation fails, MemExchange attempts insertion into an available remote slot. Only when both local and remote capacity are exhausted does the system revert to standard eviction behavior. Remote pages therefore act as overflow capacity, reducing eviction pressure on hot local items.

For \emph{GET} operations, MemExchange maintains separate hash tables for local and remote items. Because local memory offers lower access latency, the local hash table is probed first. If the item is not found locally, the remote hash table is consulted. A hit in the remote table triggers a synchronous RDMA read to retrieve the value from remote memory. A miss in both tables is reported as a cache miss. Notably, eviction decisions apply only to local items. Remote objects are excluded from local eviction selection and are treated as capacity extensions rather than as part of the primary eviction pool. This design preserves locality for frequently accessed objects while leveraging remote memory to absorb excess demand.

\subsection{Fault Tolerance}
\label{sec:fault-tolerance}

MemExchange adopts a fully distributed architecture in which each server independently manages memory exchange decisions through its local Tracker, eliminating single points of failure, localizing faults, and avoiding global coordination bottlenecks.

If a Tracker fails, only the tenants hosted on that server cease participating in future cross-node reallocations. Those tenants continue operating as standard Memcached instances using local memory only. Existing remote mappings involving other servers (and the server with the failed tracker) remain valid, and the rest of the cluster continues executing the MTC protocol without disruption.

Tenant failures are handled similarly. If a victim (donor) crashes, subsequent accesses to its remote pages are treated as cache misses by the owning victor and handled via the normal insertion path. If a victor crashes, its remote pages remain allocated in the victim’s address space but are no longer referenced. Reclaiming such orphaned pages is a potential enhancement but does not affect correctness or isolation. The reallocation mechanism tolerates transient coordination failures. Because MTC is event-driven and restartable, timeouts during victim selection trigger retries rather than deadlock. If a selected victim fails during reallocation, the victor selects another candidate. If the current highest-scored tenant in the cluster crashes, other memory-constrained tenants eventually surpass its stale score~\cite{seyri2022memsweeper}, triggering a new MTC round and resuming reallocations in the cluster. This ensures continued progress and prevents starvation.



The next section evaluates MemExchange under microbenchmark, medium-scale, and large-scale deployments to quantify its performance overheads, hit-rate improvements, and impact on cluster-wide memory utilization.

\section{Evaluation}
\label{sec:evaluation}

\subsection{Experimental Setup}
\subsubsection{System Implementation}

MemExchange is implemented on top of Memcached and consists of over 6,000 lines of modifications and additions.\footnote{Source code is available at: \href{https://github.com/AAMH/memcached}{https://github.com/AAMH/memcached}.} 
The implementation introduces cluster-wide memory sharing, shadow queues with marginal-utility scoring, the MemExchange Tracker Communication (MTC) protocol, and RDMA-based remote memory access.



\subsubsection{Testbed}

Experiments were conducted on CloudLab’s Utah cluster~\cite{Duplyakin+:ATC19} using both \emph{c6525-25g} and \emph{m510} nodes depending on compatibility requirements. The \emph{c6525-25g} nodes provide 16-core AMD EPYC CPUs, 128\,GB RAM, and Mellanox ConnectX-5 25\,Gbps NICs. The \emph{m510} nodes provide 8-core Intel Xeon D CPUs, 64\,GB RAM, and Mellanox ConnectX-3 10\,Gbps NICs.

\subsubsection{Workloads and Benchmarking Tools}

We used CloudSuite~\cite{Ferdman:173764, cloudsuite:xxx} and Mutilate~\cite{mutilate:xxx}. CloudSuite reports per-second throughput and latency statistics (average and percentiles), which we aggregate as the median across tenants during steady state. Mutilate reports summary throughput, latency, and miss rates; when aggregating across tenants, we compute weighted miss rates and median latencies for comparability.

We evaluate three workloads: \textbf{Twitter (CloudSuite)}, a skewed (Zipfian) trace with heterogeneous object sizes (80–440 bytes); \textbf{Uniform Synthetic}, a uniform workload with fixed-size (650-byte) objects; and \textbf{Facebook ETC (Mutilate)}~\cite{atikoglu2012workload}, evaluated under both uniform and Zipf access variants.



\subsection{Evaluation Methodology}

Our evaluation proceeds in four stages. We first present microbenchmarks isolating the latency cost of remote memory access (Microbenchmark~A) and the overhead of dynamic right-sizing (Microbenchmark~B). We then conduct medium-scale experiments comparing MemExchange with Memcached, MemSweeper~\cite{seyri2022memsweeper}, and Infiniswap~\cite{gu2017efficient} under Twitter and Facebook ETC workloads, measuring hit rate, memory utilization, and latency. Finally, we present a large-scale 100-node cloud experiment to evaluate scalability and convergence behavior. 
Although Memtrade~\cite{maruf2023memtrade} is architecturally related, its implementation was not publicly available at the time of evaluation; we therefore compare against MemSweeper and Infiniswap as the closest open systems.

\subsection{Microbenchmark A: Remote Memory Access Latency}
\label{sec:microA}

\subsubsection{Experimental setup and scope}

This microbenchmark quantifies the end-to-end latency of accessing remote memory through a caching layer. We measure SET and GET latency in MemExchange when items reside either in local DRAM or in remote memory exposed via different transport mechanisms. All measurements include full Memcached request processing and client round-trip time, and therefore reflect cache-level access latency rather than raw memory latency.

Each experiment involves two tenants on separate physical servers: a donor (victim) provisioned with enough local memory to store the entire working set, and a receiver (victor) provisioned with 64\,MB of local memory, with the remaining 94\% of the 1,074\,MB working set stored remotely. Both tenants receive the same synthetic uniform random workload (slab class 10) at 20K requests/s. This rate was chosen to remain below saturation for all transport schemes, ensuring that latency differences reflect transport overhead rather than overload effects.

After an initial SET-only warm-up phase, experiments run a GET-only steady-state phase in which all requests are hits. Due to the uniform access pattern and limited local memory, approximately 94\% of the victor’s GETs are served from remote memory, while the victim serves all requests locally and provides the baseline for each experiment. The victor’s performance reflects the cost of accessing data stored remotely. We evaluate three remote memory access mechanisms: \emph{RDMA} (hardware), \emph{RXE} (Software RDMA over Converged Ethernet)~\cite{man7:rxe}, and \emph{TCP} (remote instance accessed via the Memcached binary protocol). All other parameters (hardware, NIC (Mellanox ConnectX-5 25\,GB), Memcached configuration, workload, and request rate) remain identical across schemes.

\subsubsection{Local vs. remote access latency}

We first compare the victor tenant against its victim to isolate remote-access overhead.

\paragraph{GET-only phase}

Table~\ref{latency_metrics} reports the typical steady-state GET latency (median average and p99). Under the local baseline, median average latency is 22~$\mu$s with p99 of $\sim$29~$\mu$s across schemes. When serving data remotely (94\% remote portion), median average latency increases to 31~$\mu$s (41~$\mu$s p99) for RDMA, 62~$\mu$s (71~$\mu$s p99) for RXE, and 73~$\mu$s (115~$\mu$s p99) for TCP. Relative to the local baseline, remote GETs incur slowdowns of approximately $1.4\times$ (RDMA), $2.8\times$ (RXE), and $3.3\times$ (TCP). Tail amplification is most pronounced for TCP.





\paragraph{SET-only phase}

Table~\ref{latency_metrics} also reports the median average and p99 latency for SET requests. Under the local baseline, median average latency ranges from 25--26~$\mu$s, with substantially higher p99 values (557--584~$\mu$s). When inserting into remote memory, SET latency does not increase. Median average latency is 16~$\mu$s (20~$\mu$s p99) for RDMA, 26~$\mu$s (31~$\mu$s p99) for RXE, and 21~$\mu$s (25~$\mu$s p99) for TCP. In all cases, remote SET latency is comparable to or lower than the local baseline.

Unlike GET operations, which require synchronous remote reads, remote SET operations bypass Memcached’s local object insertion path and decouple local cache management from request completion. RDMA writes are issued asynchronously to remote memory and are therefore not on the critical path of SET processing. As a result, the victor experiences lower SET latency than the local baseline, which performs full in-cache insertion and metadata updates. 






\begin{table}[!htbp]
\centering
\caption{Microbenchmark A: Local vs. Remote Request Latency (Median).}
\label{latency_metrics}
\small
\begin{tabular}{|l|c|cc|cc|}
\hline
\multirow{2}{*}{\parbox{1cm}{\centering\textbf{Access Method}}} &
\multirow{2}{*}{\parbox{1cm}{\centering\textbf{Request\\Type}}} &
\multicolumn{4}{c|}{\textbf{Request Latency (\(\mu s\))}} \\
\cline{3-6}
& & \multicolumn{2}{c|}{\textbf{Remote}}
  & \multicolumn{2}{c|}{\textbf{Baseline}} \\
\cline{3-6}
& & \textbf{Avg} & \textbf{p99}
  & \textbf{Avg} & \textbf{p99} \\
\hline
\multirow{2}{*}{RDMA}
 & SET & \textbf{16}  & \textbf{20} & 25 & 557  \\
 & GET & 31  & 41 &  \textbf{22} & \textbf{30}  \\
\hline
\multirow{2}{*}{RXE}
 & SET & \textbf{26}  & \textbf{31} & 26 & 562  \\
 & GET & 62  & 71 &  \textbf{22} & \textbf{29}  \\
\hline
\multirow{2}{*}{TCP-based}
 & SET & \textbf{21}  & \textbf{25} & 26 & 584  \\
 & GET & 73 & 115 &  \textbf{22} & \textbf{29} \\
\hline
\end{tabular}
\end{table}

\subsubsection{RDMA vs. RXE vs. TCP}
\label{sec:rdmavstcp}

We next compare the three remote mechanisms directly. Across both phases, hardware RDMA consistently achieves the lowest latency, followed by RXE and then TCP. While RXE relies on software processing and UDP transport, it still outperforms the TCP-based design. Although all schemes provide functional remote access, tail behavior diverges sharply: during the GET-only phase, TCP’s p99 latency is $2.8\times$ higher than RDMA.

RDMA GET operations require synchronous remote reads: the victor cannot complete a request until the RDMA read finishes. Despite this strict synchronization requirement, hardware RDMA minimizes overhead through kernel bypass, zero-copy transfers, and NIC offload, resulting in the smallest gap relative to local access. We further optimized the completion path by replacing sleep-based polling with cooperative yielding, ensuring latency reflects RDMA service time rather than scheduling artifacts.

RXE preserves RDMA semantics but executes entirely in software over UDP, introducing additional overhead from the software RDMA stack, packet processing, and kernel involvement. This explains its consistently higher latency compared to hardware RDMA, while still outperforming TCP. In contrast, the TCP-based scheme incurs an additional full Memcached hop on the victim side, including request parsing, hash lookup, slab management, and TCP/IP stack processing. This extra indirection increases both average and tail latency. Although structurally different from RDMA-based access, it represents a common alternative for remote caching~\cite{maruf2023memtrade}, providing a meaningful comparison point. Our results show that RDMA-based approaches substantially reduce overhead relative to such TCP-based designs.

\subsection{Microbenchmark B: Resizing Overhead}

\subsubsection{Goal.}
This microbenchmark isolates the transient overhead introduced by resizing, beyond the steady-state cost of serving requests from remote memory. Specifically, it quantifies the overhead of (i) the MTC protocol and (ii) memory reallocation activity while resizing is in progress. Unlike the previous microbenchmark, which compares converged steady states, this experiment measures the incremental performance impact during active resizing.

\subsubsection{Experimental setup.}
We use two physical nodes with the CloudSuite \emph{Twitter} dataset under a Zipfian access pattern. A 1\,GB working set is offered at ~25K RPS (90\% GET, 10\% SET). The victor is provisioned with 64\,MB local memory, with the remainder supplied via MemExchange until its effective capacity reaches 1\,GB, after which resizing is disabled. We collect application-level metrics (throughput, average latency, p90–p99) and system-level measurements (CPU utilization and network counters, including UDP and RDMA statistics). Due to skewed access, most requests are served locally even after resizing; RDMA counters confirm that steady-state serving traffic remains limited. This setup isolates resizing overhead while keeping the request-serving path largely unchanged.

\paragraph{Quiet point detection and window selection.}
To separate resizing from steady state, we define a \emph{quiet point} as the first timestamp at which UDP traffic drops to zero and remains zero. Since MTC coordination uses UDP, zero UDP deltas indicate that resizing has stopped. RDMA traffic is excluded from detection because it remains active in steady state. We analyze two 60-second windows: an \emph{active} window preceding the quiet point and a \emph{quiet} window following it. These windows isolate resizing overhead while minimizing long-term drift.

\begin{table}[!htbp]
\centering
\caption{Microbenchmark B: Resizing overhead during active and quiet windows (60\,s each).}
\label{resize_overhead}
\small
\resizebox{\columnwidth}{!}{%
\begin{tabular}{lcccc}
\hline
\textbf{Metric} & \multicolumn{2}{c}{\textbf{Victor}} & \multicolumn{2}{c}{\textbf{Victim}} \\
                & \textbf{Active} & \textbf{Quiet} & \textbf{Active} & \textbf{Quiet} \\
\hline
\multicolumn{5}{l}{\emph{Request latency}} \\
Avg latency ($\mu$s)            & 26.9  & 23.9  & 22.8  & 22.4  \\
p99 latency ($\mu$s)            & 86.0  & 48.3  & 34.0  & 31.0  \\
\hline
\multicolumn{5}{l}{\emph{CPU utilization}} \\
CPU utilization (\%)            & 10.46  & 9.68  & 6.75  & 6.73  \\
Load (1\,min)                   & 3.44   & 3.19  & 2.23  & 2.17  \\
\hline
\multicolumn{5}{l}{\emph{MTC control traffic (UDP)}} \\
UDP packets/s                   & 5.32   & 0.43  & 5.39  & 0.40 \\
UDP bytes/s                     & 326.7  & 0     & 319.2 & 0 \\
\hline
\multicolumn{5}{l}{\emph{RDMA NIC activity (RoCE)}} \\
RDMA recv data (KB/s)           & 2194   & 2234  & 1288  & 108   \\
RDMA xmit data (KB/s)           & 1309   & 109   & 2176  & 2211  \\
RDMA read reqs/s (target-side)  & 0      & 0     & 1390  & 1447  \\
RDMA write reqs/s (target-side) & 0      & 0     & 807   & 4     \\
\hline
\end{tabular}%
}
\end{table}

\subsubsection{Latency and throughput overhead.}

Table~\ref{resize_overhead} compares application-level performance during active and quiet windows. On the victor, throughput remains stable at ~25K RPS, indicating that resizing does not throttle request admission. From the active to the quiet window, average latency decreases from 26.9\,$\mu$s to 23.9\,$\mu$s and p99 latency decreases from 86\,$\mu$s to 48.3\,$\mu$s, corresponding to resizing overheads of ~3\,$\mu$s (average) and 37.7\,$\mu$s (p99).

On the victim, latency changes are minimal and slightly improve after resizing completes (22.8\,$\mu$s to 22.4\,$\mu$s average; 34\,$\mu$s to 31\,$\mu$s p99), confirming that resizing mainly affects the victor.

\paragraph{Remote GET stability.}
Because remote GETs synchronously wait for RDMA completion, latency differences could reflect either resizing overhead or changes in remote-hit frequency. However, victim-side RDMA read counters remain stable at ~1400 reads/s across both windows (Table~\ref{resize_overhead}), indicating that the fraction of remote GETs does not change. The latency reduction after the quiet point therefore reflects the removal of resizing-related control-plane and reallocation activity rather than changes in steady-state serving behavior.



\subsubsection{CPU overhead.}
During the active window, the victor’s CPU utilization is 10.46\%, decreasing to 9.68\% in the quiet window, reflecting the additional work of resizing logic. The victim shows a similarly small change (6.75\% to 6.73\%). Load averages decline from 3.44 to 3.19 (victor) and from 2.23 to 2.17 (victim), indicating that resizing introduces only modest and transient CPU overhead.



\subsubsection{Network and RDMA activity.}
Network counters clearly separate resizing from steady state.
During the active window, the MTC protocol generates sustained UDP control traffic (~5 packets/s, 320--330 bytes/s per node), which drops to near zero once resizing completes. RDMA data-plane activity shows a complementary transition.
On the victor, received RDMA traffic remains stable at ~2.2\,MB/s across both windows, while transmit traffic decreases from ~1.3\,MB/s during resizing to ~0.1\,MB/s afterward.
The victim mirrors this pattern: transmit traffic stays near ~2.2\,MB/s, whereas received traffic drops from ~1.3\,MB/s to ~0.1\,MB/s.

Victim's RDMA request counters clarify the cause. Write requests fall from ~807/s during resizing to near zero once resizing stops, while read requests remain steady at ~1.4K/s. The disappearance of writes reflects the end of remote page allocation and associated \emph{SET} activity, whereas reads persist due to steady-state \emph{GET} servicing. Together, these measurements show that resizing introduces transient UDP control traffic and RDMA writes tied to page allocation.
After convergence, control traffic and write-heavy reallocation cease, while steady-state RDMA reads remain unchanged.

Overall, resizing introduces only transient control-plane traffic and additional RDMA writes associated with page allocation. Once convergence is reached, latency, throughput, and RDMA read behavior return to steady-state levels, indicating that resizing overhead is confined to the active resizing interval.

\subsection{Medium-scale Benchmarks}
\label{sec:mediumscalebenchmarks}

\subsubsection{Twitter (Zipfian)}

We compare MemExchange against Memcached, MemSweeper~\cite{seyri2022memsweeper}, and Infiniswap~\cite{gu2017efficient} on a 10-server cluster. Because Infiniswap requires kernel 4.4, experiments in this section are conducted on CloudLab \emph{m510} nodes (Mellanox ConnectX-3 10\,Gbps NICs). While absolute performance differs from newer hardware, all systems are evaluated under identical conditions.

Each server hosts two tenants. Three servers contain only over-provisioned tenants, three contain only under-provisioned tenants, and four host a mix of both. Over-provisioned tenants are allocated 2\,GB of local memory; under-provisioned tenants have only 64\,MB. All tenants receive a 1\,GB dataset driven by a CloudSuite client at 25K requests/s (25 connections, 90:10 GET:SET). Keys and values follow the skewed distribution of the Twitter trace. After a warm-up phase that preloads the dataset, systems enter main phase, during which additional memory is dynamically allocated according to each system’s mechanism. MemExchange may allocate remote memory when local memory is insufficient, whereas MemSweeper relies solely on local reallocation. Because the workload is highly skewed, the majority of requests target hot items that remain in local memory, while remote accesses primarily serve colder items.

Behavioral differences are most evident on servers hosting only under-provisioned tenants. On mixed servers, over-provisioned tenants reduce memory pressure and schemes behave similarly. On under-provisioned-only servers, MemExchange allocates remote memory while MemSweeper remains locally constrained, resulting in fundamentally different handling of overflowed data.

\paragraph{Infiniswap}
\label{sec:infiniswap-intro}

Unlike MemExchange, Infiniswap~\cite{gu2017efficient} does not perform workload-aware right-sizing. Instead, it preserves each tenant’s configured memory allocation and transparently swaps pages to remote memory when local DRAM is insufficient.

In our experiments, Infiniswap-backed Memcached tenants are provisioned with a logical memory allocation at least as large as their working set, but their local DRAM usage is capped via \texttt{cgroups} to match the local-memory constraints imposed on MemExchange victors. This configuration forces a comparable fraction of the working set to reside in remote swap, ensuring a fair comparison of remote-access overhead. When local DRAM limit is exceeded, the operating system swaps pages to remote memory via Infiniswap rather than to disk.

From Memcached’s perspective, the entire key space always fits in memory and no items are evicted; all accesses are cache hits served either from local DRAM or remote swap. Consequently, Infiniswap tenants exhibit an effective 100\% cache hit rate. Performance differences therefore reflect the latency overhead of serving a portion of hits from remote swap rather than local memory.

Next, we compare systems in terms of hit rate, memory utilization, and request latency. Latency is reported as the median of per-second measurements across tenants, with tail behavior evaluated using the median per-second p99, capturing typical per-tenant experience while preserving sensitivity to tail amplification.

\paragraph{Hit Rate}

We evaluate hit rate across 10 under-provisioned (victor) tenants; over-provisioned tenants maintain 100\% hit rate under all systems and are excluded. As shown in Table~\ref{twitter_combined}, all schemes begin with nearly identical initial hit rates due to identical allocations. Memcached reaches a final hit rate of 86.9\%, improving through natural warm-up but remaining constrained by static memory limits. MemSweeper increases the final hit rate to 92.0\%, corresponding to a 38.6\% miss-rate reduction relative to Memcached. Gains arise primarily on mixed servers, where under-provisioned tenants reclaim unused local memory from co-located victims.

MemExchange further improves the final hit rate to 95.2\%, reducing miss rate by 63.1\% relative to Memcached. The additional benefit over MemSweeper stems from under-provisioned-only servers, where MemExchange allocates remote memory to extend effective working-set capacity beyond local DRAM limits. Infiniswap maintains a structural 100\% hit rate because all data remains resident via remote swap. While this eliminates misses, it incurs higher latency due to page-fault and block-layer overhead. Overall, MemExchange achieves the highest effective working-set coverage among adaptive systems without relying on swap-based indirection.


\paragraph{Memory Utilization}

We next examine cluster-wide memory redistribution. Initially, utilization is approximately 50\%: 10 under-provisioned tenants have 64\,MB each, while 10 over-provisioned tenants have 2\,GB each, leaving substantial unused capacity.

MemExchange reclaims and redistributes approximately 10\,GB to under-provisioned tenants, effectively doubling their useful working-set capacity. Cluster-wide utilization increases by roughly 50\%, directly enabling the discussed hit-rate gains. Reallocations occur both on mixed servers and on under-provisioned-only servers by leveraging remote memory to eliminate stranded capacity. MemSweeper improves utilization by approximately 20\%. Because it reallocates only locally, gains are limited to mixed servers; homogeneous under-provisioned servers remain memory-constrained. Infiniswap similarly increases effective memory via remote swap, yielding aggregate utilization comparable to MemExchange. However, it does not perform workload–aware right-sizing; instead, all pages remain resident via remote swap, eliminating misses at the cost of higher latency, as discussed next.

\paragraph{Request Latency}

Table~\ref{twitter_combined} reports the median per-second average and p99 latency across tenants. Memcached achieves the lowest latency, as all accesses are local, but also the lowest hit rate. MemSweeper incurs modest overhead from periodic local reallocation while avoiding RDMA costs, keeping latency close to baseline.

MemExchange introduces slightly higher latency due to remote accesses. Microbenchmark~A~(\S\ref{sec:microA}) showed that a fully remote GET incurs an incremental penalty of ~9~$\mu$s (31~$\mu$s vs.\ 22~$\mu$s median average). Under the skewed Twitter workload, receiver-side RDMA counters show approximately ~1,900 remote GETs per second out of ~22,500 total GETs ($\sim$8–9\%). Applying a 9~$\mu$s penalty to this fraction predicts an average increase of <1~$\mu$s, consistent with measured results. Thus, most hot items remain local, and remote accesses primarily serve colder overflow data. Tail effects are more visible: because ~8–9\% of requests incur higher remote latency, these disproportionately influence the top 1\% of the distribution, leading to a modest p99 increase relative to MemSweeper.

SET operations do not materially affect latency. Remote SETs are not on the critical path and constitute only 10\% of requests in this workload. Infiniswap exhibits substantially higher average and tail latency. Although it maintains 100\% hit rate, remote swap incurs page-fault handling and block-layer overheads beyond direct RDMA access, resulting in roughly $5.5\times$ higher average latency and nearly $13.8\times$ higher p99 latency than Memcached.

\begin{table}[!htbp]
\centering
\small
\setlength{\tabcolsep}{3pt}
\renewcommand{\arraystretch}{1.0}
\begin{tabular}{lcccc}
\toprule
\textbf{Metric} & \textbf{Memcached} & \textbf{MemSweep.} & \textbf{\name} & \textbf{Infini.} \\
\midrule
Avg lat. & 36.2~$\mu$s & 43.4~$\mu$s & 43.6~$\mu$s & 201~$\mu$s \\
p99 lat. & 109~$\mu$s  & 299~$\mu$s  & 319~$\mu$s  & 1500~$\mu$s \\
\midrule
Final HR  & 86.9\% & 92.0\% & 95.2\% & 100\% \\
Miss Red. & 0\%    & 38.6\% & 63.1\% & 100\% \\
\bottomrule
\end{tabular}
\caption{Twitter workload. Initial hit rates $\sim$77\% across schemes. Miss reduction relative to Memcached. Latencies are median per-second values across tenants.}
\label{twitter_combined}
\end{table}

\subsubsection{Facebook ETC (Zipfian)}

We evaluate MemExchange against Memcached, MemSweeper~\cite{seyri2022memsweeper}, and Infiniswap~\cite{gu2017efficient} using the Facebook ETC workload~\cite{atikoglu2012workload} on the same 10-server cluster described earlier. To maintain comparability with Infiniswap, all experiments in this section run on \emph{m510} nodes.

Each server hosts two tenants (three over-provisioned-only, three under-provisioned-only, and four mixed). Over-provisioned tenants are allocated 2\,GB of local memory; under-provisioned tenants have 64\,MB. Each tenant serves a 1\,GB dataset generated via \texttt{mutilate} using Facebook ETC key and value size distributions, driven at 25K requests/s (25 connections, 95:05 GET:SET). After a sequential warm-up phase, systems dynamically adjust memory according to their respective mechanisms.

We evaluate two ETC variants: \textbf{Uniform ETC}, the default mutilate configuration with uniformly sampled keys; and \textbf{Zipf ETC}, where keys follow a Zipf distribution ($\alpha=1.1$) to introduce strong popularity skew consistent with prior analyses~\cite{atikoglu2012workload}. The two ETC variants represent distinct locality regimes. Uniform ETC provides limited cache locality and higher miss pressure, whereas Zipf ETC concentrates requests on hot keys that remain resident in local memory, reducing miss rates and sustaining higher effective throughput. Together, they allow evaluation under both low- and high-locality conditions. As in the Twitter evaluation, we report hit rate, memory utilization, and request latency (median and p99 across tenants during steady state).

\paragraph{Overall Hit Rate}

Table~\ref{etc_results} shows distinct behavior under Uniform and Zipf ETC due to their different locality characteristics.

Under Uniform ETC, accesses are evenly distributed across the working set, yielding limited locality. With a 64\,MB cache serving a 1\,GB dataset, Memcached exhibits a high miss rate. Memory reallocation remains beneficial: MemSweeper reduces misses on mixed servers via local reclamation, while MemExchange further improves hit rate by allocating remote memory on other servers. Infiniswap reports zero misses by utilizing remote swap.

Under Zipf ETC ($\alpha=1.1$), locality is stronger and Memcached’s miss rate drops to 11.6\%, indicating that a small hot set captures most requests. MemSweeper reduces misses to 10.9\%, and MemExchange further to 9.8\%. Although the absolute improvement appears modest, at 25K RPS per tenant, a 1.8\% reduction translates to thousands of avoided backend fetches per second cluster-wide.

As in the Twitter workload, differences between MemSweeper and MemExchange are visible on under-provisioned-only servers. On mixed servers, both reclaim idle local memory; on homogeneous under-provisioned servers, only MemExchange can allocate remote memory. Under both ETC variants, this additional capacity allows MemExchange to retain a larger fraction of the working set. Overall, ETC results confirm that MemExchange improves hit rate both when locality is weak (Uniform) and when skew is present (Zipf), consistently outperforming purely local reallocation.


\begin{table}[!htbp]
\centering
\small
\setlength{\tabcolsep}{2pt}
\renewcommand{\arraystretch}{1.0}
\resizebox{\columnwidth}{!}{
\begin{tabular}{llccccc}
\toprule
\textbf{Workload} & \textbf{Scheme} 
& \textbf{Miss \%} 
& \textbf{R Avg} 
& \textbf{R p99} 
& \textbf{U Avg} 
& \textbf{U p99} \\
\midrule

\multirow{4}{*}{Uniform}
& Memcached   & 94.4 & 53.0 & 107 & 63.2 & 122 \\
& MemSweeper  & 83.5 & 66.4 & 120 & 80.4 & 136 \\
& MemExchange & 68.7 & 74.7 & 131 & 85.4 & 150 \\
& Infiniswap  & 0    & 416  & 1573 & 606  & 1998 \\
\midrule

\multirow{4}{*}{Zipf ($\alpha=1.1$)}
& Memcached   & 11.6 & 50.0 & 86.2 & 55.2 & 92.4 \\
& MemSweeper  & 10.9 & 53.6 & 108  & 59.8 & 116  \\
& MemExchange & 9.8  & 67.4 & 179  & 73.4 & 189  \\
& Infiniswap  & 0    & 103  & 329  & 139  & 476  \\
\bottomrule
\end{tabular}
}
\caption{ETC workload under Uniform and Zipf access. Miss rate weighted by requests; latencies in $\mu$s (median across tenants). R and U denote read and update latencies, respectively.}
\label{etc_results}
\end{table}

\begin{figure*}[!htbp]
\centering
\includegraphics[width=\textwidth, height=7cm]{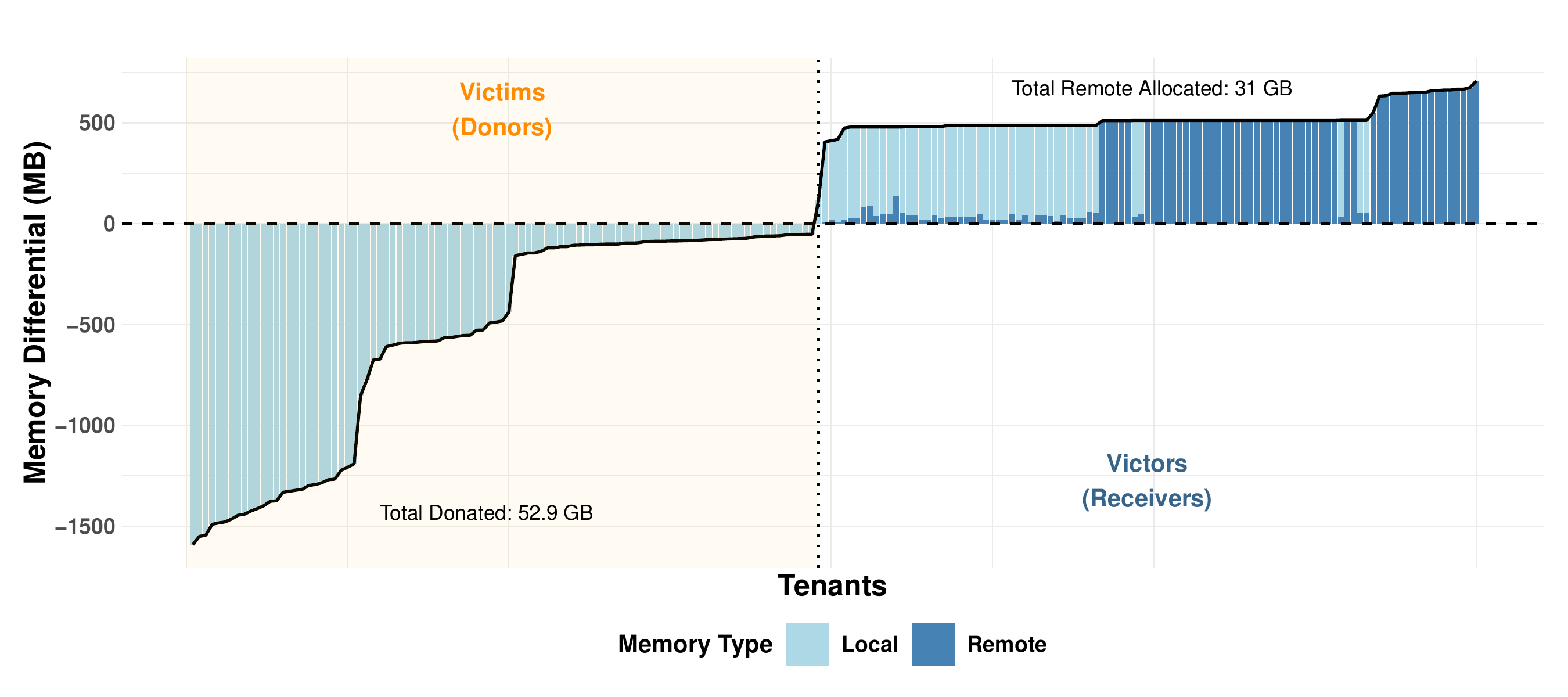}
\caption{Cluster-wide memory redistribution across 200 tenants, sorted by total memory differential (local + remote). Negative values denote over-provisioned tenants relinquishing memory; positive values denote under-provisioned tenants receiving capacity. Light blue bars show local adjustments and dark blue bars remote RDMA allocations.}
\label{fig10}
\end{figure*}

\paragraph{Memory Utilization}

Memory utilization under ETC is independent of the access distribution (Uniform vs.\ Zipf), as dataset size, tenant allocations, and reallocation mechanisms are identical. Each tenant serves a 1\,GB dataset; under-provisioned tenants have 64\,MB of local memory and over-provisioned tenants 2\,GB, resulting in $\sim$50\% initial cluster utilization. MemExchange reclaims unused memory from over-provisioned tenants and redistributes it cluster-wide, including to homogeneous under-provisioned servers via remote allocation. This eliminates nearly all stranded capacity and improves utilization by $\sim$50\%, consistent with the Twitter workload.

MemSweeper also reclaims idle memory but is limited to local reallocation. Stranded capacity on under-provisioned-only servers remains unused, leading to a smaller overall improvement. Infiniswap achieves a similar aggregate utilization to MemExchange by keeping entire working sets resident via remote swap. However, it does not prioritize reclaiming idle local memory on mixed servers, as it assumes tenants are already right-sized and that memory pressure arises from insufficient server-level main memory. While effective in raising utilization, this approach introduces the latency overhead discussed next.

\paragraph{Request Latencies}

Table~\ref{etc_results} reports median per-tenant latencies under Uniform and Zipf ETC. Unlike Twitter, ETC uses a heavy-tailed Pareto inter-arrival distribution, leading to burstier arrivals and visible queuing effects.

Under Uniform ETC, elevated miss rates increase backend pressure. Baseline Memcached sustains 10.9K QPS with 53~$\mu$s median read latency (107~$\mu$s p99). MemSweeper and MemExchange incur additional overhead from reallocation and, for MemExchange, remote accesses; MemExchange reports 74.7~$\mu$s median read latency (131~$\mu$s p99). Because uniform access lacks a sharply defined hot set, remote requests occur more frequently, making their penalty more visible. MemExchange therefore trades moderate latency increases for substantial miss-rate reduction. Infiniswap performs significantly worse (416~$\mu$s median, $>$1.5~ms p99), as swap-backed accesses trigger page faults and block-layer overhead, which amplify queuing under bursty load.


Under Zipf ETC ($\alpha=1.1$), strong skew concentrates requests on a small hot set and all systems sustain 25K QPS. Memcached achieves 50~$\mu$s median read latency (86.2~$\mu$s p99); MemSweeper remains close at 53.6~$\mu$s (108~$\mu$s p99), while MemExchange reports 67.4~$\mu$s (179~$\mu$s p99). The modest average-latency increase for MemExchange reflects the fraction of synchronous remote GETs, which lie on the critical path. As in Microbenchmark~A(\S\ref{sec:microA}), remote accesses primarily affect tail latency, explaining the larger p99 gap. Infiniswap improves relative to Uniform (103~$\mu$s median) due to stronger locality but remains slower than all other systems.


\subsection{Large-scale Benchmark}
\label{sec:largescalebenchmark}

To evaluate MemExchange under realistic cloud-scale conditions, we conducted a large-scale experiment spanning 100 servers (CloudLab \emph{c6525-25g}), deploying 200 tenants with two tenants co-located per node. Each server was provisioned with \SI{4}{GB} of shared memory, evenly divided between its two resident tenants. Tenants were randomly designated as either \emph{under-provisioned} (victors) or \emph{over-provisioned} (victims), modeling the heterogeneous and imbalanced memory demands commonly observed in production clusters.

Under-provisioned tenants were assigned working sets of \SI{2.5}{GB} but constrained to \SI{2}{GB} of local memory, deliberately inducing sustained memory pressure. Over-provisioned tenants remained idle and acted as potential donors of unused memory. Memory-hungry tenants requested additional capacity via the MTC protocol, and MemExchange redistributed memory across the cluster according to its utility-driven reallocation policy. This decentralized coordination continuously rebalances supply and demand while preserving locality and avoiding centralized bottlenecks.

Each tenant was driven at 25\,K requests per second using a GET-dominant workload derived from the CloudSuite Twitter dataset. The access pattern followed a sequential progression through the working set, creating sharp performance cliffs once local capacity was exceeded. This workload is particularly sensitive to memory shortfalls and therefore well suited for stress-testing cluster-wide reallocation mechanisms.

The experiment ran continuously for 12 hours. Memory reallocations converged after approximately 7 hours, after which the system operated in a stable steady state under sustained load. This convergence period reflects the gradual redistribution of memory across 100 servers, as MemExchange incrementally reallocates capacity based on measured utility while avoiding disruptive oscillations.

Over the course of the experiment, MemExchange reallocated a total of \SI{53}{GB} of memory cluster-wide, including \SI{32}{GB} served remotely via RDMA. This redistribution increased overall memory utilization by 13.25\%, demonstrating MemExchange’s ability to reclaim and repurpose idle capacity at scale. Among the 104 under-provisioned tenants, 99 achieved sustained 100\% hit rates, and 4 additional tenants exceeded 90\%, indicating that MemExchange effectively eliminates performance cliffs for the vast majority of memory-constrained workloads even in large deployments.

\begin{figure}[!htbp]
\centering
\includegraphics[width=0.45\textwidth, height=9cm]{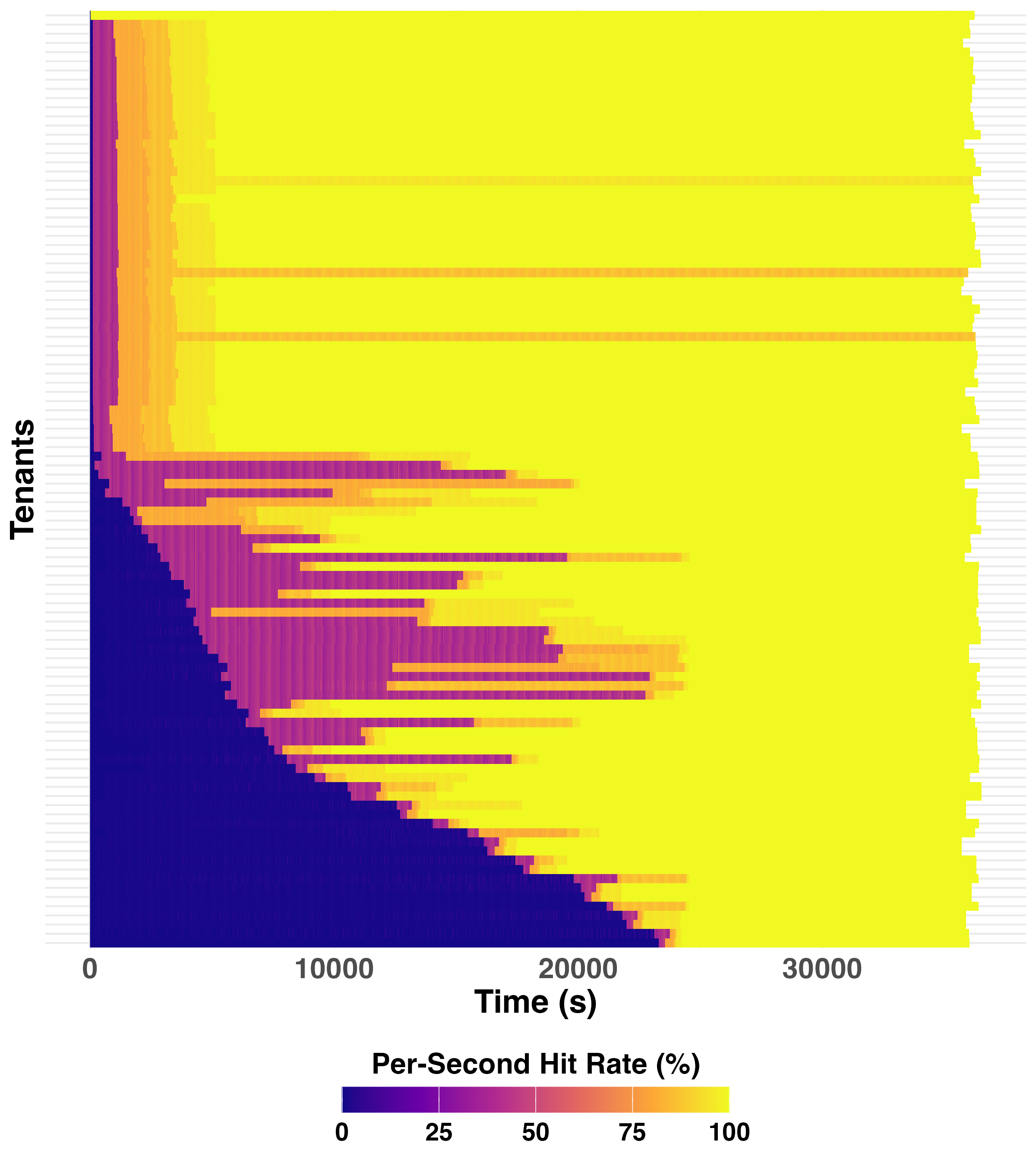}
\caption{Per-second hit rate heatmap for under-provisioned tenants during the large-scale experiment.}
\label{heatmap}
\end{figure}

Figure~\ref{fig10} visualizes the final memory differentials across all 200 tenants, highlighting the asymmetric but coordinated redistribution of capacity between victors and victims. 
To better expose convergence behavior, tenants are sorted by the 25th percentile of their per-second hit rate, capturing the duration and severity of low-hit-rate intervals during memory acquisition. As illustrated in Figure~\ref{heatmap}, tenants with prolonged near-zero hit rates appear toward the bottom, while faster-converging tenants cluster near the top, revealing a clear gradient of stabilization across the cluster. 




\section{Conclusion}
\label{sec:conclusion}

In this paper, we presented MemExchange, a cluster-scale memory management system driven by marginal utility for multi-tenant in-memory caching. MemExchange dynamically redistributes idle memory across physical nodes using RDMA, improving both individual tenant hit rates and overall cluster memory utilization without requiring careful tenant placement or co-location.

To enable efficient remote memory access and low-overhead reallocation, we introduced the MemExchange Tracker Communication (MTC) protocol, which coordinates victim selection and memory transfer while avoiding remote CPU involvement in the data path. This design preserves fast local access for hot items while using remote memory as an overflow tier to relieve capacity pressure.

Across medium-scale and rack-scale deployments—including a 100-server experiment with 200 tenants—MemExchange improved memory utilization by up to 50\% in medium-scale benchmarks and reallocated up to \SI{53}{GB} cluster-wide, increasing overall utilization by 13.25\% at rack scale. Compared to TCP-based remote memory approaches, RDMA reduced request latency by up to $3.5\times$ while maintaining stable convergence at scale. These results demonstrate that fine-grained, marginal-utility–guided memory reallocation is both practical and effective at cluster scale. By transforming memory from a static per-tenant allocation into a dynamically shared resource, MemExchange provides a viable foundation for efficient memory disaggregation and memory-as-a-service deployments.
\section{Discussion and Future Work}
\label{sec:future}

MemExchange leaves several avenues for future improvement.

First, remote memory currently serves primarily as an overflow tier and does not actively migrate items back to local memory. Under long-running or highly skewed workloads, frequently accessed remote items may incur sustained RDMA overhead. Adaptive migration policies (such as promoting hot remote items or evicting persistently cold ones) could further improve latency and throughput. Item size–aware placement may also reduce steady-state RDMA costs for large items, prioritizing their placement in local memory when remote fetch latency would be disproportionately high.

Second, remote pages are not reclaimed once allocated, which simplifies bookkeeping but may limit long-term flexibility under tenant churn. Mechanisms for recycling or rebalancing remote pages would improve memory efficiency and fault tolerance.

Third, while the MTC protocol’s wait-for-score mechanism favors stability over responsiveness, proactive victim selection based on historical scoring or predictive heuristics could reduce reallocation delay while preserving decision quality.

Finally, while MemExchange targets caching systems where utility is expressed as hit rate, the marginal-utility framework may extend to other distributed or heterogeneous memory environments where resource benefit can be quantified.

\bibliography{refer.bib}



\end{document}